\documentclass[12pt,twoside]{article}   
\usepackage[super,sort,comma]{natbib}

\usepackage{amsmath}
\usepackage{hyphenat}
\usepackage{multirow}
\usepackage{fancyhdr}		
\usepackage{showkeys}		

\usepackage[section]{placeins}   %
\usepackage{booktabs}
\usepackage{graphicx}
\usepackage{subcaption}
\usepackage{amsmath}   
\usepackage{amsfonts}

\makeatletter \renewcommand\@biblabel[1]{$^{#1}$} \makeatother
\newcommand{\cen}[1]{\begin{center} #1 \end{center}}

\usepackage{hyperref}
\hypersetup{ colorlinks,
    citecolor=blue,
    filecolor=blue,
    linkcolor=blue,
    urlcolor=blue
}

\usepackage{xcolor}

\definecolor{gray}{rgb}{0.6,0.6,0.6}
\definecolor{red}{rgb}{0.85,0,0}
\definecolor{green}{rgb}{0,0.85,0}
\definecolor{blue}{rgb}{0,0,0.85}
\definecolor{beige}{rgb}{0.92,0.87,0.78}
\usepackage[all]{hypcap}    
\usepackage{bm}

\begin{document}

\cen{\sf
  {\Large {\bfseries Biomechanical Mapping of Tumor Growth: A Novel Method to Quantify Glioma Infiltration and Mass Effect}}\\[10mm]

  Carles López-Mateu\textsuperscript{1,*},
  Maria Gómez-Mahiques\textsuperscript{1},
  F.\,Javier Gil-Terrón\textsuperscript{1},
  Víctor Montosa-i-Micó\textsuperscript{1},\\
  Donatas Sederevičius\textsuperscript{2},
  Kyrre E.\,Emblem\textsuperscript{2},
  Juan M.\,García-Gómez\textsuperscript{1},
  Elies Fuster-García\textsuperscript{1}\\[4mm]

  \textsuperscript{1}Biomedical Data Science Laboratory, Instituto Universitario de Tecnologías de la Información y Comunicaciones,\\
  \hspace*{2em}Universitat Politècnica de València, 46022 Valencia, Spain\\[1mm]
  \textsuperscript{2}Department of Physics and Computational Radiology, Division of Radiology and Nuclear Medicine,\\
  \hspace*{2em}Oslo University Hospital, 0372 Oslo, Norway\\[5mm]

  Version typeset \today\\
}

\pagenumbering{roman}
\setcounter{page}{1}
\pagestyle{plain}


Email: \href{mailto:carleslopezmateu@proton.me}{clopmat@upv.es}

\begin{abstract}
\noindent {\bf Background:}Glioblastoma (GBM) growth can alter surrounding brain tissue through location-dependent physiological changes. Two main growth phenotypes—(I) infiltrative, marked by diffuse invasion with minimal mass effect, and (II) proliferative, marked by pronounced tissue compression—are recognized, yet their quantitative characterization and prognostic impact remain poorly explored.\\[4pt]

{\bf Purpose:} To develop and validate an MRI-based biomarker, the  \emph{Dynamic Infiltration Rate} (DIR), that quantifies the balance between tumor volume expansion and peritumoral compression, and to evaluate its ability to stratify patients by overall survival (OS).\\[4pt]

{\bf Methods:} The DIR was defined as the ratio between tumor‐volume enlargement and mass-effect–induced peritumoral compression. We performed technical validation using a synthetically generated dataset with known ground truth spanning realistic infiltrative–proliferative spectra. Patients were dichotomized into high- and low-infiltration groups by threshold that maximizes the log-rank statistic for OS; multivariate Cox regression adjusted for age, sex, and MGMT status.\\[4pt]

\textbf{Results:} In silico validation showed high concordance with ground truth ($R^2 = 0.85$). Clinically, low DIR was associated with significantly improved overall survival (median = 35.2 weeks) compared to high DIR (median = 16.0 weeks; $p = 0.0001$). DIR effectively stratified patients by patient survival (log-rank p < 0.001, HR = 2.49), and remained an independent prognostic factor on multivariate analysis (HR = 1.38, 95\% CI 1.12–1.70; p = 0.0027).\\[4pt]

{\bf Conclusions:} The DIR is a novel, robust quantitative MRI biomarker that distinguishes proliferative from infiltrative phenotypes in GBM and independently predicts OS. Early identification of these phenotypes could guide personalized treatment strategies and follow-up schedules.\\
\end{abstract}


\tableofcontents

\newpage

\setlength{\baselineskip}{0.7cm}      

\pagenumbering{arabic}
\setcounter{page}{1}
\pagestyle{fancy}
\section{Introduction}

Glioblastoma (GBM) is one of the hardest cancers to treat. It grows fast, relapses after initial treatment, and shows high intra-tumoral heterogeneity~\cite{comba2021uncovering}. Consequently, the outlook for people diagnosed with a GBM is poor~\cite{mirimanoff}. Identifying prognostic factors, including age, incidence rate of seizures, extent of surgical resection, and clinical performance status, has facilitated the development of crucial prognostic indices to guide patient management~\cite{council}. 

Of these factors, tumor growth in GBM is intricately linked to the mass effect, which is characterized by increased intracranial pressure, displacement of surrounding brain structures, and potential brain herniation \cite{brain_herniation}. Mechanical stress not only induces clinical symptoms, but may also contribute to neuronal dysfunction~\cite{seano}. In addition, the resultant hypoxia and hypoperfusion, caused by local dysregulation of blood vessels, promote invasiveness of cancer cells \cite{Jain}\cite{drumm}. This invasiveness, along with hindered delivery of therapeutic agents to affected regions, creates a significant obstacle to successful treatment, underscoring the critical impact of tumor infiltration on disease progression.  

The existing literature relies on metrics that assess the mass effect of tumor growth through gross displacement measures (e.g., midline shift \cite{midline} and lateral ventricle displacement~\cite{steed}). However, these metrics do not capture direct tumour behaviour in terms of its invasiveness, leaving us with an incomplete understanding of infiltration-driven tumor progression.

To this end, our study introduces a novel MRI-based approach to quantify infiltration through localized tissue compression analysis. Although existing methods rely on displacement fields to assess the mass effect~\cite{leow2007statistical}, our method uses Jacobian maps to detect subtle volumetric changes invisible to conventional techniques~\cite{paper_elies}. Specifically, we hypothesize that infiltrative phenotypes will exhibit lower peritumoral compression for equivalent increases in tumor volume compared to proliferative phenotypes.  

To test this hypothesis, we introduce the 'Dynamic Infiltration Rate (DIR)' as a measure that compares how much the tumor has expanded (growth) with overall peritumoral compression (derived from tissue compression maps). This ratio helps distinguish more infiltrative tumors - characterized by relatively low compression for a given amount of growth - from those that are more proliferative, where compression is higher.

In our experiments, we validate the DIR using an in-house synthetic tumor model that simulates both infiltration and proliferation. Next, we iteratively derive an optimal threshold to categorize patients from two longitudinal datasets (Lumière and UCSF) and perform survival analyses. Finally, we used multiparametric Cox regression to compare DIR with established prognostic factors, including MGMT status, age, and sex.

\section{Materials and Methods}
\subsection{Synthetic tumour growth model}

We developed a lightweight, fully‑parametric simulator that
\emph{jointly} reproduces (i) the \textbf{mass effect} caused by
tumour proliferation and (ii) the \textbf{contrast halo} produced by
diffuse cell infiltration (see Appendix \ref{appendix:tumor_growth_model}.  Starting from a baseline
contrast‑enhanced T1‑weighted image \(I_0\) and its tumour‑core mask
\(M\), the model requires only two scalar inputs:

\begin{itemize}
\item \textbf{Proliferation gain \(\alpha\) (mm).}  
      Each core voxel seeds an anisotropic Gaussian potential; summation and scaling by \(\alpha\) give the growth kernel \(K\!\), whose gradient yields a displacement field 
      \[
        \widetilde{\bm{u}} = \nabla K.
      \]
      Here, \emph{displacement field} means a vector-valued function defined at every point in the image, whose direction and length indicate both where and by how much that small volume of tissue is shifted. 
      We normalise the field so that
      \(\max\lVert\widetilde{\bm{u}}\rVert = \alpha\) mm, i.e.\ the numeric value of \(\alpha\) equals the intended peak tissue shift.  Clinical brain-shift magnitudes of 0–5 mm  (rarely \(\sim10\) mm) \cite{gerard2017brain} are therefore reproduced with \(\alpha=0\!-\!5000\).  Tissue-stiffness modulation, Perlin heterogeneity and boundary attenuation are applied exactly as detailed in Appendix A, and the warped image is denoted \(I_\alpha\).
\item \textbf{Infiltration extent \(\phi\) (voxels).}
      Pure invasion is simulated by dilating the core mask
      \(\phi\) times with a 1‑voxel sphere,
      \(M_\phi=M\oplus^\phi B\).  
      Newly reached voxels copy an intensity from a
      \(3{\times}3{\times}3\) neighbourhood in the original core and
      are finally blurred to yield a smooth contrast halo.  
      Because no geometry changes, \(\phi\) is in centimeters (cm).
\end{itemize}

By sweeping \(\alpha\in[0,5000]\) mm and
\(\phi\in[0,8]\) vox (0–8 cm), we generate the family of synthetic
images \(I_{\alpha,\phi}\), spanning realistic combinations of
mechanical displacement and diffuse infiltration.
A full mathematical derivation, implementation notes and parameter
tables are provided in Appendix~A.

\subsection{Clinical Datasets}

Two clinical datasets were used. The first, the Lumiere Dataset \cite{suter2022lumiere}, comprises 91 patients with Glioblastoma who underwent surgical resection followed by temozolomide‑based chemoradiation at the University Hospital of Bern between 2008 and 2013. This dataset is longitudinal—with patients evaluated at an average of approximately 7 timepoints—allowing for a more nuanced assessment of disease progression. For voxel‑wise annotations, Lumiere provides automated segmentation outputs generated by two state‑of‑the‑art tools.

The second, the University of California San Francisco (UCSF) Dataset \cite{california}, is a publicly available, annotated collection of multi‑modal brain MRI scans from 298 patients with diffuse glioma, each evaluated at two consecutive follow‑up visits (596 scans in total). This dataset includes expert voxel‑wise annotations delineating four key tumor subregions: enhancing tissue (ET), surrounding non‑enhancing FLAIR hyperintensity (SNFH), non‑enhancing tumor core (NETC), and resection cavity (RC). It also provides longitudinal annotations detailing volumetric changes in the ET and SNFH compartments between the two follow‑up scans.

\subsection{Patient Selection Criteria}
The inclusion criteria for both datasets were: (1) an increase in enhancing tumor volume, (2) progressive disease according to RANO criteria, and (3) availability of overall survival (OS) and IDH-status data. Only patients with IDH-wildtype tumors were included, since the 2021 WHO Classification of Tumours of the Central Nervous System reserves the term glioblastoma for IDH-wildtype lesions \cite{WHO}. For the Lumiere dataset, T1-contrast-enhanced (T1c), T2-weighted, and fluid-attenuated inversion recovery (FLAIR) MRI modalities were mandatory for in-house tumor segmentation; in contrast, the UCSF dataset already provided the segmentations.

\subsection{Quantification of Tissue Displacement, Compression, and Tumor Growth}

Tissue--displacement maps were computed for the synthetic data set following the strategy described in~\cite{paper_elies}. Beginning with the baseline T1c image ($\alpha = 0$, $\phi = 0$), this reference image was non-linearly registered to every other synthetic image generated for the different $(\alpha,\phi)$ combinations using the \textit{Greedy: Fast Deformable Registration for 3D Medical Images} algorithm~\cite{greedy,greedy_git}. The registration used normalized cross-correlation with a $4\times4\times4$ neighborhood as the similarity metric, a metric-gradient regularization of $1.723$~voxels, a metric-warp regularization of $0.707$~voxels, and maximum warp-field precision. The gradient-descent step size was $0.25$. The optimizer ran for $200$, $150$, and $20$ iterations at the three resolution levels, with shrink factors of $4$, $2$, and $1$, respectively.

As the tumor evolves, cancer cells infiltrate the surrounding tissue, leading to intensity changes in T1c images. Such changes may be erroneously interpreted by the registration algorithm as spatial shifts, resulting in inaccurate displacement field calculations. To address this issue, the tumor core—comprising both enhancing tumor and necrotic tissue—was excluded from the registration process via a mask. This exclusion ensures that the displacement fields are computed solely based on non-tumoral tissue within the peritumoral area, thereby avoiding artifacts caused by intensity variations.

The Jacobian Map (JM) is obtained by taking the logarithm of the determinant of the Jacobian matrix, which quantifies local volumetric changes in the tissue. For a given displacement field \(\mathbf{T}(x,y,z) = \big( T_x, T_y, T_z \big)\), the Jacobian matrix is defined as:
\[
\mathbf{J}(x,y,z) = \begin{pmatrix}
\frac{\partial T_x}{\partial x} & \frac{\partial T_x}{\partial y} & \frac{\partial T_x}{\partial z} \\
\frac{\partial T_y}{\partial x} & \frac{\partial T_y}{\partial y} & \frac{\partial T_y}{\partial z} \\
\frac{\partial T_z}{\partial x} & \frac{\partial T_z}{\partial y} & \frac{\partial T_z}{\partial z}
\end{pmatrix}.
\]

Thus, the JM is computed as
\[
\mathrm{JM}(x,y,z)=\log\!\bigl(\det\mathbf{J}(x,y,z)\bigr).
\]

Next, TCM (Tissue Compression Map) is computed as
\[
\mathrm{TCM}(x,y,z)=\max\!\bigl(0,\,-\mathrm{JM}(x,y,z)\bigr).
\]

In our study, the displacement fields are defined on a discrete voxel grid, and spatial derivatives are approximated using finite differences between neighboring voxels. To quantify tissue compression induced by tumor growth, the median TCM value within the peritumoral region of interest (ROI) was computed.

Tumor growth was assessed using the segmented tumor mask. To mitigate non-linear scaling effects inherent to volumetric measurements, the tumor core volume was converted into an equivalent spherical radius. This equivalent radius represents the radius of a sphere that would have the same volume as the tumor core. The change in tumor size between consecutive time points was then determined by computing the difference between the equivalent spherical radii at those time points.

\begin{figure}
    \centering
    \includegraphics[width=0.8\linewidth]{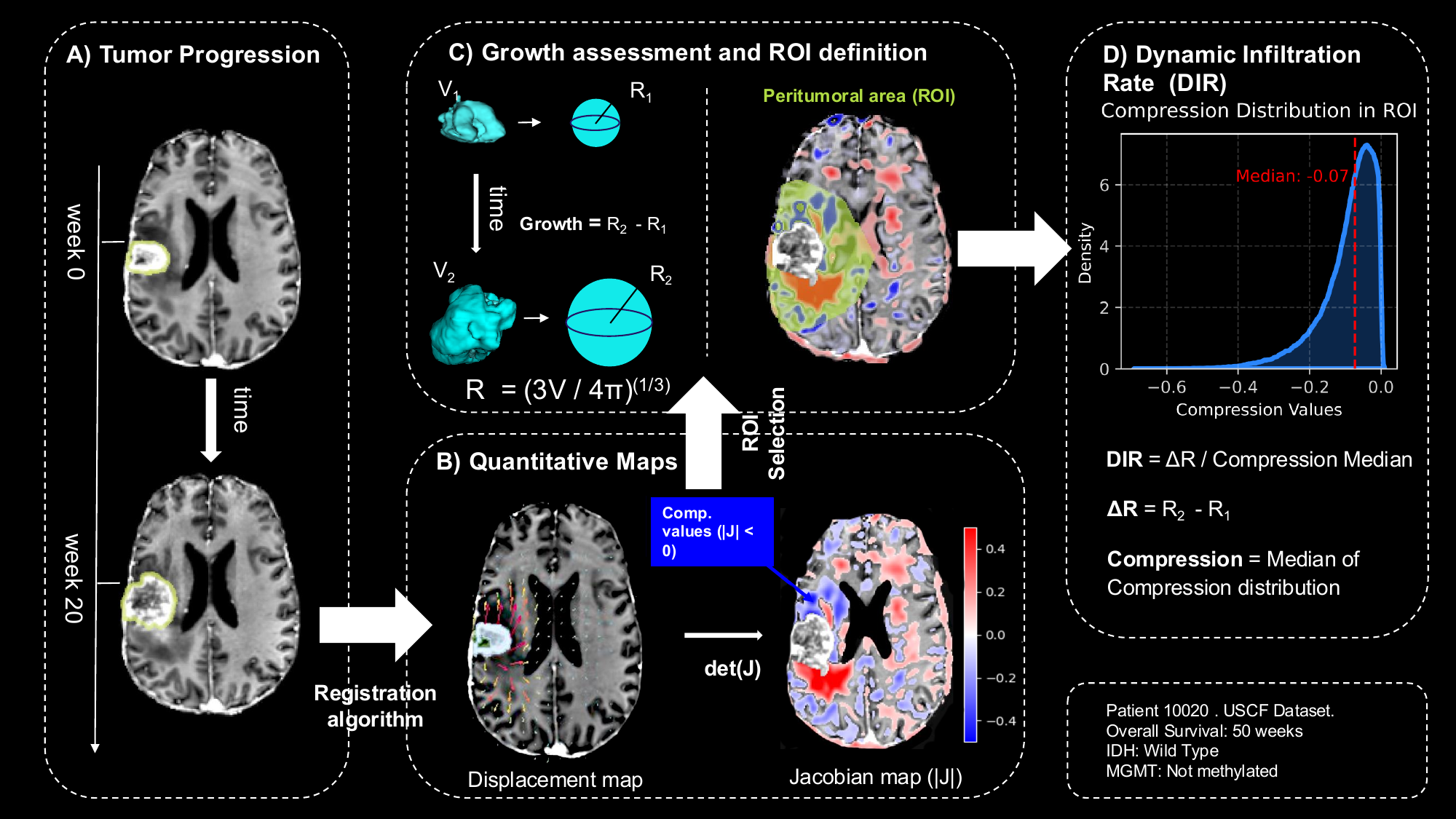}
    \caption[Workflow for DIR computation]%
  {%
  \textbf{Workflow for computing the Dynamic Infiltration Rate (DIR).}  
  (A) Baseline (week 0) and follow-up (week 20) contrast-enhanced T1-weighted MRIs illustrate tumour progression.  
  (B) Non-linear registration between time-points yields a voxel-wise displacement field whose logarithmic Jacobian determinant produces a tissue-compression map (TCM).  
  (C) The peritumoral region of interest (PRI) is generated by dilating the tumour core mask by 3 cm and subtracting the core; growth is quantified as the change in equivalent spherical radius (\(\Delta R\)).  
  (D) The \(\mathrm{DIR}= \Delta R / \mathrm{median}\{\mathrm{TCM}_{\text{PRI}}\}\) normalises growth by surrounding compression, providing a size-independent index of infiltration.  
  Patient metadata are shown for context.%
  }
  \label{fig:workflow}
\end{figure}

Figure \ref{fig:workflow} illustrates the workflow of the methodology.

\subsection{Dynamic Infiltration Rate}

The \textit{Dynamic Infiltration Rate} (DIR) is a composite metric that quantifies the relationship between tumor growth and the mechanical compression exerted on the surrounding tissue. In this study, two variants of the DIR are defined, corresponding to the experimental approach and the simulation-based ground truth.

\subsubsection*{Experimental DIR}
For the clinical and synthetic datasets, tumor growth is quantified as the change in the equivalent spherical radius of the tumor core (\(\Delta R_{\text{core}}\)), derived from the tumor segmentation. Tissue compression is measured using the TCM, by computing the median compression values within the peritumoral region of interest ($Med(TCM_{PRI})$). The PRI is delineated by dilating the tumor core mask by 3 cm and subtracting the core region, thereby including only the tissue affected by the mass effect. Thus, the experimental DIR is expressed as:
\[
\text{DIR} = \frac{\Delta R_{\text{core}}}{Med(JM_{PRI})}.
\]

\subsubsection*{Ground Truth DIR}
In the synthetic tumor growth simulations, the ground truth DIR is computed from the known contributions of proliferation and infiltration. In this context, \(\Delta R_{\text{prolif}}\) represents the increase in the tumor radius attributable solely to proliferative growth, while \(\Delta R_{\text{inf}}\) corresponds to the increase derived exclusively from infiltration. Therefore, the ground truth DIR is defined as:
\[
\text{DIR} = \frac{\Delta R_{\text{inf}}}{\Delta R_{\text{prolif}}}.
\]

Comparing these two measures enables an assessment of the experimental approach's ability to capture the true dynamics of tumor infiltration, as determined by the simulation parameters.

\subsection{Survival Analysis}

The fundamental step of our analysis involves \text{binary stratification of tumor phenotypes} by identifying an optimal threshold that directly correlates with patient survival data. This threshold is estimated using log rank test complemented by kaplan-meier curves. In instances of censored data, we assign the date of the last known patient contact or the date of the most recent MRI examination as the censorship date, whichever is applicable. In addition, we calculate the average values of the DIR for individuals with multiple observations to ensure robust statistical evaluations.

Subsequently, a multiparametric Cox proportional hazards model was used to assess the impact of various co-factors on patient survival. This model incorporates demographic information available in the datasets, including gender and age, as well as molecular characteristics such as MGMT methylation status~\cite{mgmt}. 

Statistical analyzes were performed in two distinct phases. Each data set was initially analyzed separately: the \textbf{Lumiere Dataset} and the \textbf{UCSF Dataset}. Subsequently, a combined analysis was performed using data from both datasets.

\section{Results}

\subsection{Patient Population: final cohort}
\begin{figure}
\centering
\includegraphics[scale=0.]{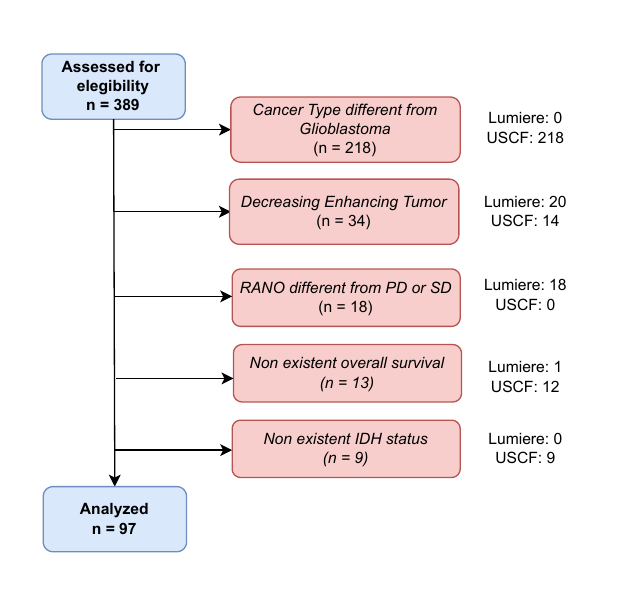}
\caption[CONSORT flow diagram]%
  {%
  \textbf{CONSORT flow diagram for cohort assembly.}  
  From 389 screened glioblastoma cases (Lumière \(n=91\), UCSF \(n=298\)), we applied longitudinal-imaging availability, RANO progression, overall-survival data, and IDH-wild-type status criteria, arriving at a final study cohort of 97 patients (Lumière \(n=44\); UCSF \(n=53\)). Boxes indicate exclusion reasons and the remaining sample size at each step.%
  }
\label{fig:CONSORT flow diagram}
\end{figure}

After applying the inclusion criteria (see Figure \ref{fig:CONSORT flow diagram}), a total of 97 patients remained in the final cohort, comprising 53 from the USCF dataset and 44 from the Lumiere Dataset. In USCF dataset the mean age was 60 years (SD = 10.7), with 31 males and 22 females; regarding MGMT status, 0 patients were methylated, 37 were unmethylated, and 3 were missing. In contrast, in the Lumiere Dataset, the mean age was 57.6 years (SD = 9.0), with 22 males and 22 females; for qualitative MGMT status, 10 patients were unmethylated, 12 were methylated, and 9 had missing data.

\subsection{Experimental DIR Assessment}
\begin{figure}
    \centering
    \includegraphics[width=0.5\linewidth]{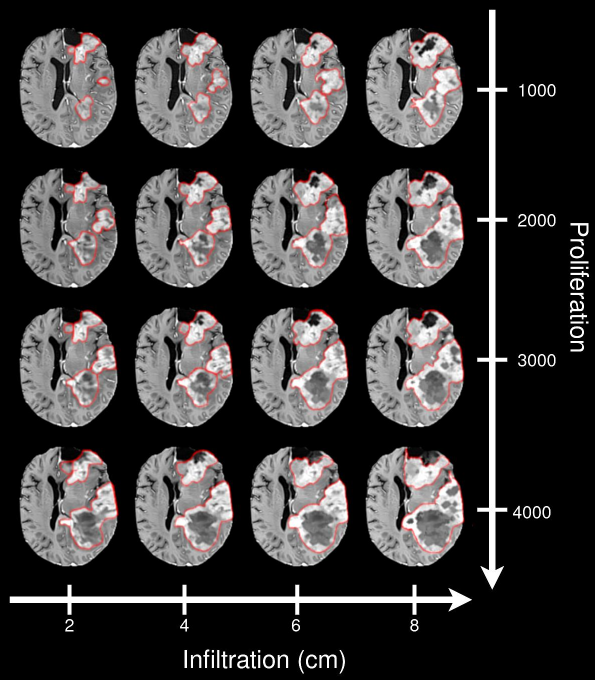}
    \caption{\textbf{Patient Simulation Example}. Each row in the figure corresponds to a distinct \(\alpha\) value, representing different levels of proliferative activity. In contrast, each column corresponds to a specific \(\phi\) value, indicating varying degrees of tissue infiltration.} 
    \label{fig:Synthetic-growth grid}
\end{figure}

Using the tumor growth model developed, we generated 2,000 synthetic growth images from 10 baseline images of 10 different patients from Lumiere dataset. The tumor core volumes in all baseline patient images exhibited a mean volume of \(46.95\, \text{cm}^3\) with a standard deviation of \(40.68\, \text{cm}^3\). Figure~\ref{fig:Synthetic-growth grid} presents representative examples of the synthetic images generated by varying \(\alpha\) and \(\phi\), from a single baseline image.

\begin{figure}
    \centering
    \includegraphics[width=0.5\linewidth]{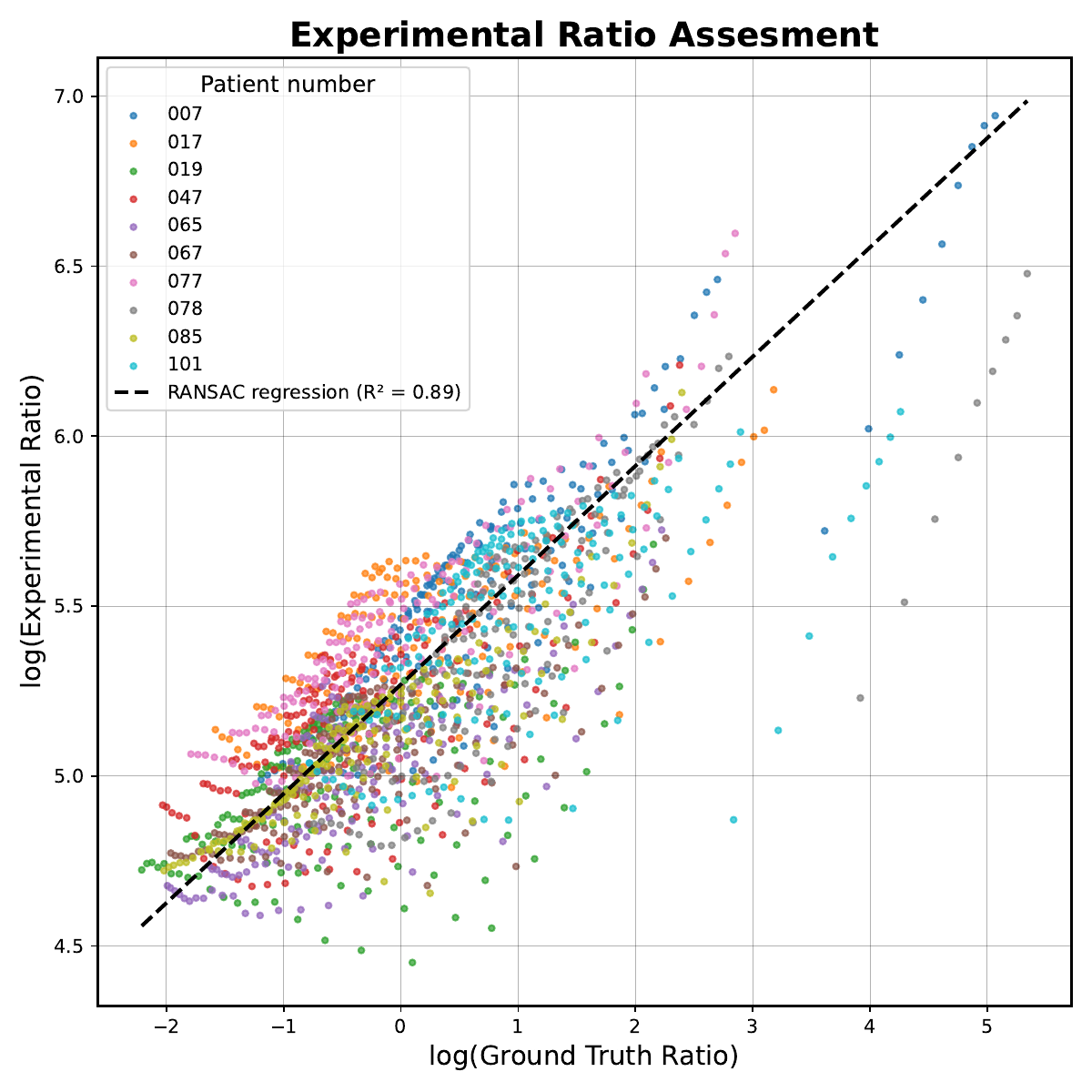}
    \caption[DIR validation plot]%
  {%
  \textbf{Validation of the experimental DIR against ground-truth simulations.}  
  Log–log scatter plot of experimental DIR versus the known proliferation-to-infiltration ratio for 2 000 synthetic images generated from 10 baseline patients..%
  }
    \label{fig:DIR validation plot}
\end{figure}

The experimental DIR was compared with the ground truth DIR (Figure \ref{fig:DIR validation plot}), producing an R2 in the RANSAC regression of 0.89.

\subsection{Survival Analysis: application to longitudinal datasets}
\subsubsection{Stratification Based on DIR}
\begin{figure}[htbp]                 
  \centering

  \begin{subfigure}[b]{0.48\textwidth}
    \includegraphics[width=\linewidth]{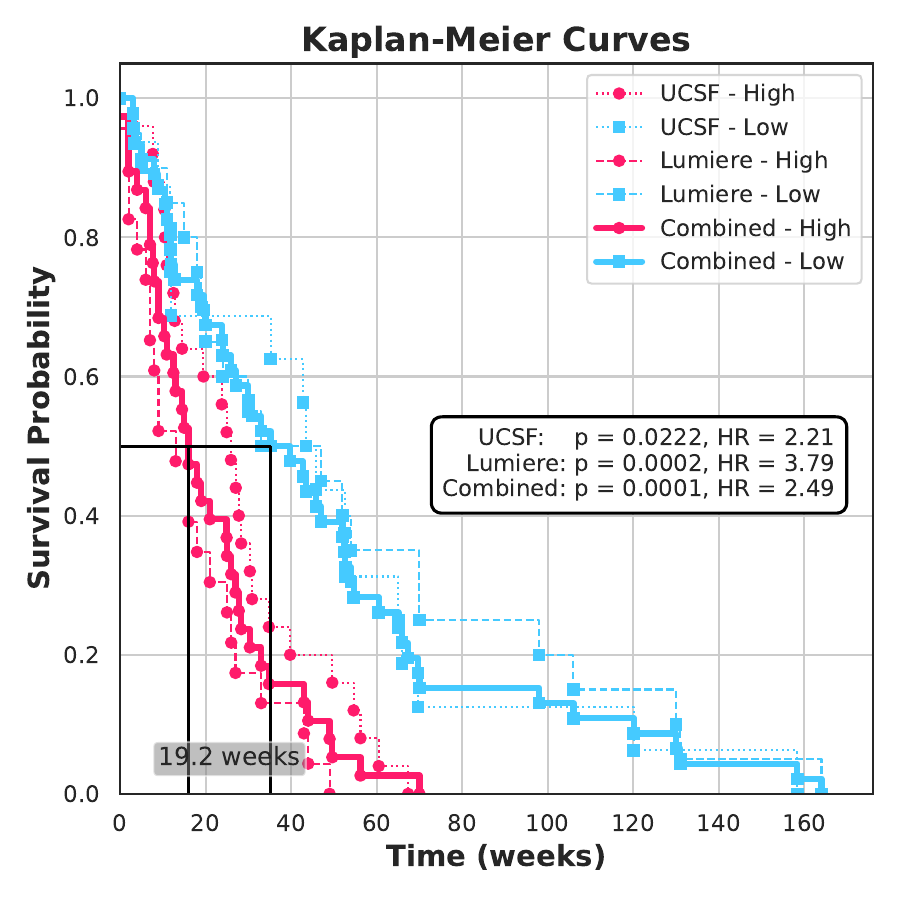}
    \caption{}                      
    \label{fig:kaplan-meier}                  
  \end{subfigure}\hfill
  \begin{subfigure}[b]{0.48\textwidth}
    \includegraphics[width=\linewidth]{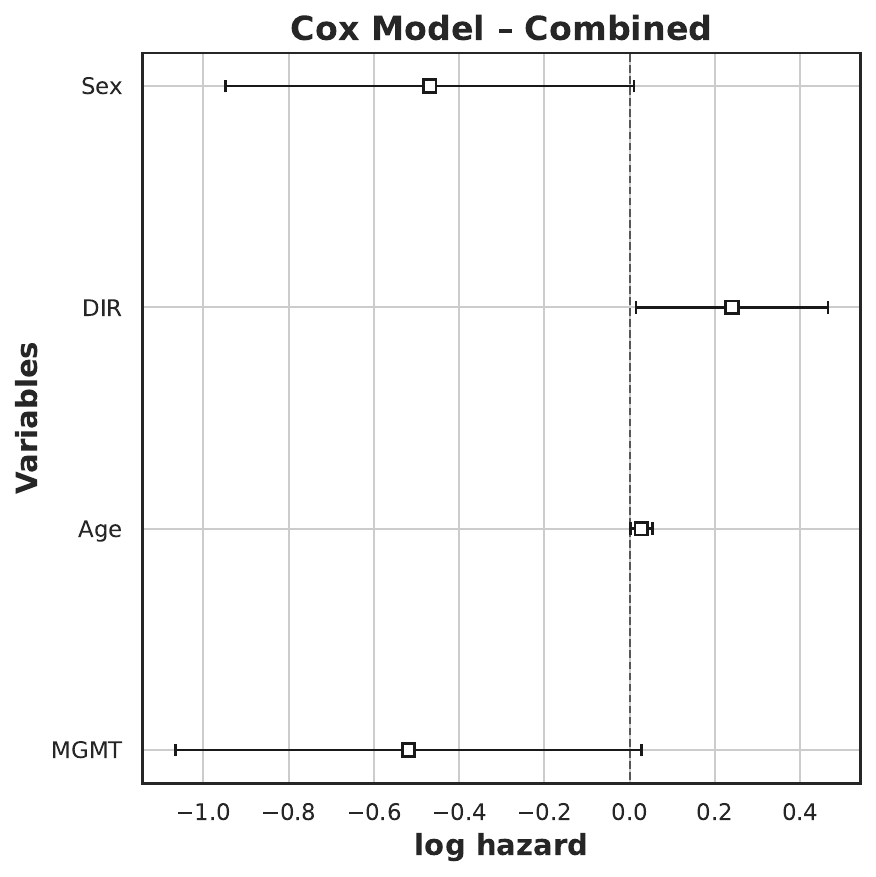}
    \caption{}                      
    \label{fig:cox_combined}
  \end{subfigure}

  \vspace{0.5em} 

  \begin{subfigure}[b]{0.48\textwidth}
    \includegraphics[width=\linewidth]{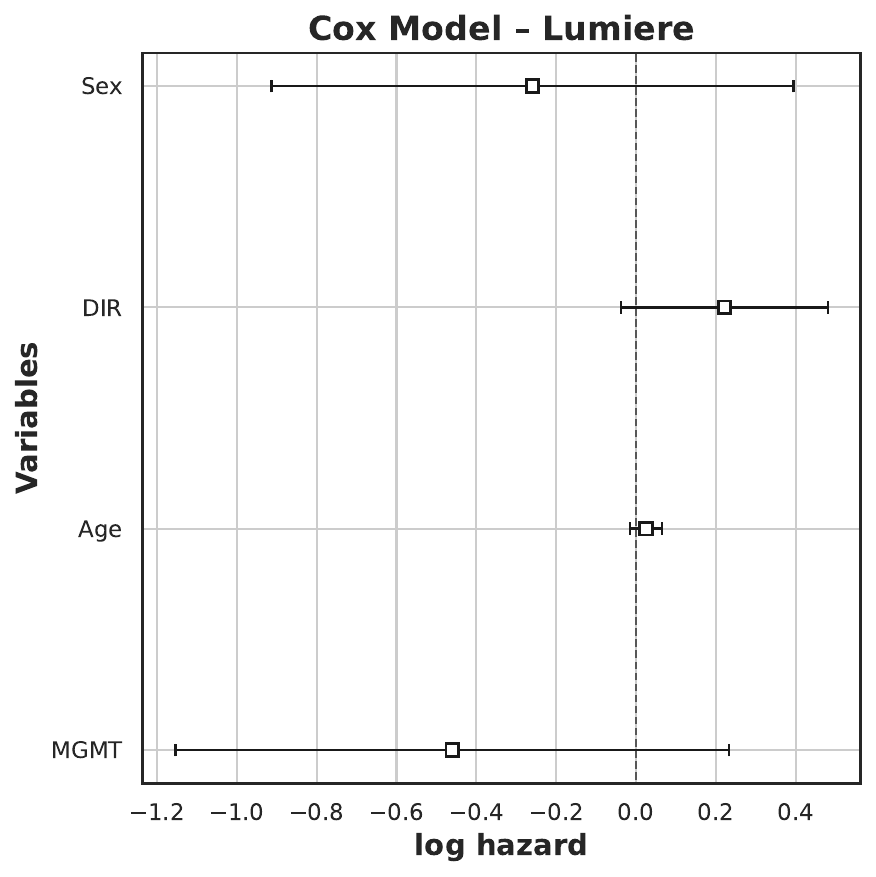}
    \caption{}                      
    \label{fig:cox_lumiere}
  \end{subfigure}\hfill
  \begin{subfigure}[b]{0.48\textwidth}
    \includegraphics[width=\linewidth]{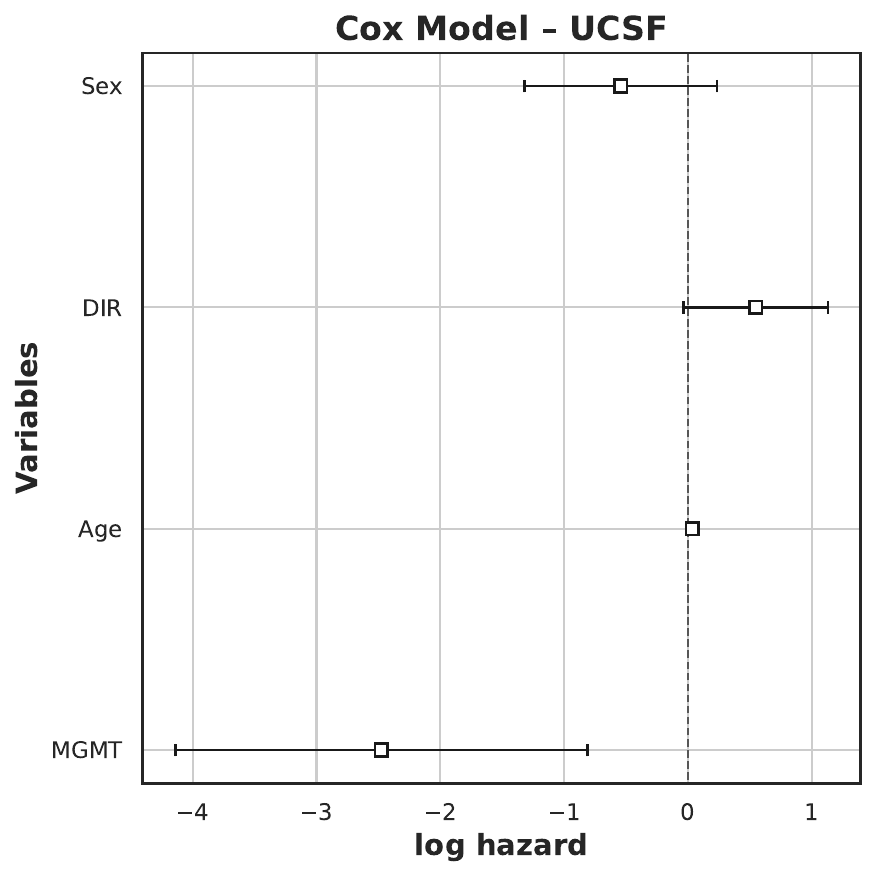}
    \caption{}                      
    \label{fig:cox_USCF}
  \end{subfigure}

  \caption[Kaplan–Meier and Cox models]{%
    \textbf{Prognostic power of the Dynamic Infiltration Rate (DIR).}  
    (a) Kaplan–Meier curves show significantly longer overall survival for low-DIR tumours in the Lumière, UCSF and combined cohorts.  
    (b–d) Forest plots of multivariate Cox models adjusted for age, sex and MGMT status reveal high DIR as an independent adverse factor in the combined dataset (b) and in each cohort ((c) Lumière, (d) UCSF).%
  }
  \label{fig:km_cox_grid}
\end{figure}

In the Lumière cohort, an optimal threshold (DIR*) of 79.90~mm was identified, yielding a hazard ratio (HR) of 3.79 and $p=0.000219$, with the difference between the medians (Low–High) amounting to 32.0~weeks, indicating significantly longer median survival in the low-DIR group; in the UCSF cohort, an optimal threshold of 33.54~mm was determined with HR=2.21 and $p=0.022224$, resulting in a median difference of 17.5~weeks in favor of the low-DIR group; and in the combined dataset, an optimal threshold of 66.09~mm was established with HR=2.49 and $p=0.000105$, with the median difference being 19.2~weeks, again demonstrating superior median survival for the low-DIR group. These findings are summarized in Table~\ref{tab:results_datasets}, and the corresponding Kaplan–Meier curves are presented in Figure~\ref{fig:kaplan-meier}(a).

\setlength{\tabcolsep}{12pt}%
\begin{table}[ht]
\centering
{\small
\begin{tabular}{lcccc}
\toprule
\textbf{Cohort}    & \textbf{DIR* (mm)} & \textbf{HR} & \textbf{$p$-value} & \textbf{MSD (weeks)} \\
\midrule
Lumière            & 79.90            & 3.79        & 0.000219           & 32.0                \\
UCSF               & 33.54            & 2.21        & 0.022224           & 17.5                \\
Combined           & 66.09            & 2.49        & 0.000105           & 19.2                \\
\bottomrule
\end{tabular}}
\caption{Optimal cutpoints for DIR by cohort, associated hazard ratios, significance levels, and median survival differences (Low–High Infiltration).}
\label{tab:results_datasets}
\end{table}

\subsubsection{Multivariate Cox Proportional Hazards Analysis}
We conducted multivariate Cox regression to quantify the prognostic impact of DIR in the context of clinical covariates (age, sex, MGMT methylation) for each cohort. Model concordance indices (C-index) were 0.70 (Lumière), 0.64 (UCSF) and 0.67 (combined), indicating good discriminative ability.

\paragraph{Lumière cohort (n = 44).}
Within the Lumière series, DIR showed a borderline association with survival (coefficient = 0.221; HR = 1.2475, 95 \% CI 0.9624–1.6171; $p=0.0949$), suggesting a 25\% increase in hazard per standard-deviation rise. Age (HR = 1.0250 per year; $p=0.2269$), female sex (HR = 0.7714; $p=0.4366$) and MGMT unmethylation (HR = 0.6310; $p=0.1930$) were not significant.

\paragraph{UCSF cohort (n = 53).}
In UCSF, DIR again trended toward significance (coefficient = 0.5475; HR = 1.7289, 95\% CI 0.9653–3.0965; $p=0.0656$). MGMT unmethylation emerged as a strong adverse factor (HR = 0.0841 for methylated vs unmethylated; $p=0.00354$). Age (HR = 1.0345; $p=0.0630$) and sex (HR = 0.5800; $p=0.1701$) did not reach significance.

\paragraph{Combined cohort (n = 97).}
In the pooled dataset, DIR was a significant independent predictor (coefficient = 0.2404; HR = 1.2712, 95\% CI 1.0147–1.5939; $p=0.0369$), conferring a 27\% increased hazard per SD. Age was also prognostic (HR = 1.0278; $p=0.0360$), female sex showed a trend toward protection (HR = 0.6254; $p=0.0549$), and MGMT status was marginal (HR = 0.5951; $p=0.0629$).

\paragraph{Integration of the DIR composite metric.}
Across cohorts, inclusion of DIR improved model fit markedly and rendered its individual components non-significant (all $p>0.20$), demonstrating that DIR encapsulates the prognostic information of diffusion and mass-effect measures into a single robust biomarker.

\begin{table}[ht]
  \centering
  {\scriptsize
  \resizebox{\textwidth}{!}{%
    \begin{tabular}{lcccccc}
      \toprule
      \multirow{2}{*}{\textbf{Covariate}}
        & \multicolumn{2}{c}{\textbf{Lumière (n=44)}}
        & \multicolumn{2}{c}{\textbf{UCSF (n=53)}}
        & \multicolumn{2}{c}{\textbf{Combined (n=97)}}\\
      \cmidrule(lr){2-3}\cmidrule(lr){4-5}\cmidrule(lr){6-7}
        & HR (95\% CI) & $p$
        & HR (95\% CI) & $p$
        & HR (95\% CI) & $p$\\
      \midrule
      DIR (per SD)          & 1.25 (0.96–1.62) & 0.0949 & 1.73 (0.97–3.10) & 0.0656 & 1.27 (1.01–1.59) & 0.0369\\
      Age (per year)        & 1.03 (0.98–1.07) & 0.2269 & 1.03 (1.00–1.07) & 0.0630 & 1.03 (1.00–1.05) & 0.0360\\
      Female sex            & 0.77 (0.40–1.48) & 0.4366 & 0.58 (0.27–1.26) & 0.1701 & 0.63 (0.39–1.01) & 0.0549\\
      MGMT unmethylated     & 0.63 (0.32–1.26) & 0.1930 & 0.08 (0.02–0.44) & 0.0035 & 0.60 (0.34–1.03) & 0.0629\\
      \bottomrule
    \end{tabular}%
  }}
  \caption{Multivariate Cox regression results for DIR and clinical covariates in the Lumière, UCSF, and combined cohorts (n=97).}
  \label{tab:cox_multivar}
\end{table}

~\section{Conclusion and discussion}
The complex interaction between GBM growth and its impact on adjacent brain tissue remains a critical area of research in neurooncology. The aggressive infiltration of the tumor not only suggests a poor prognosis due to the increased risk of recurrences, but also presents a challenge in understanding the subtleties of its behavior. Although macroscopic evaluations of brain tissue dynamics, such as ventricular displacements~\cite{steed} and midline shifts~\cite{gamburg}, provide insight into morphological alterations induced by tumor growth, a detailed analysis distinguishing between infiltrative and proliferative tumor characteristics is still lacking. Our study aims to close this gap by introducing a novel biomarker to accurately characterize tumor infiltration based on the analysis of longitudinal magnetic resonance imaging studies.

We observed a strong correlation between the DIR, a measure of tumor invasion, and patient survival rates, indicating that a higher DIR correlates with poorer outcomes. This underscores the aggressive nature of more infiltrative tumors~\cite{alfonso} and provides a quantitative basis to assess the growth behavior of GBM in vivo. 

Furthermore, applying multiparametric Cox regression analysis allowed a detailed examination of survival predictors, including the DIR, alongside established prognostic markers such as MGMT methylation status and patient demographics. The DIR emerged as a relevant independent predictor of survival, underscoring its potential as a valuable clinical tool for prognosis and treatment planning. 

However, our methodology, in particular the use of Jacobian maps to distinguish between invasive and proliferative tumor growth, still encounters limitations; for example, the change in intensity due to infiltration of edema into new areas during the development of tumour growth results in fake expansions and compressions. Future research should explore registration methods that are invariant to changes in intensity.

Despite these challenges, our study, supported by widely accessible tools and a straightforward analysis, demonstrates the existence of tumor growth phenotypes and its relationship with OS. This underscores the importance of further investigating the biomechanics of GBM and its impact on disease progression.

\section{Funding}
This work was funded by Grant PID2021-127110OA-I00 (PROGRESS) funded by MCIN/AEI/10.13039/501100011033 and by the European Regional Development Fund (ERDF), SINUÉ (INNEST/2022/87 – Agència Valenciana de la Innovació, Spain), Programa INVESTIGO (INVEST/2022/298) and Grant CIAICO/2022/064 (Lalaby-Glio) funded by Generalitat Valenciana, Spain. Funding was also received from the Spanish Ministry of Science, Innovation and Universities (FPU23/02431), the European Research Council (grant 758657-ImPRESS), the Helse Sør-Øst Regional Health Authority (grants 2021057 and 2017073), and the Research Council of Norway (grants 325971, 261984, 303249, and 32543). This work was further supported by the Universitat Politècnica de València through its program of pre-doctoral contracts for doctoral training (PAID-01-24, Subprogram 1).

\section{Conflicts of Interest}
The authors report no conflicts of interest.

\clearpage

\section*{Appendix}
\addcontentsline{toc}{section}{\numberline{}Appendix}
\section{Tumor Growth Model}
\label{appendix:tumor_growth_model}

In this work, the displacement field representing tumor growth is derived mathematically by combining Gaussian contributions from each tumor voxel. Let \(\Omega \subset \mathbb{R}^3\) be the image domain, and let \(x \in \Omega\) denote the coordinates of a given voxel. Consider a tumor mask \(T \subset \Omega\), where \(T\) is the set of voxels belonging to the tumor.

Each voxel \(x_t \in T\) acts as a source of local growth influence modeled by a Gaussian function:
\[
w_{x_t}(x) = \exp\left(-\frac{\|x - x_t\|^2}{2\sigma^2}\right),
\]
where \(\sigma > 0\) controls how rapidly the influence of the tumor voxel decreases with distance.

Summing the contributions from all tumor voxels, we have:
\[
S(x) = \sum_{x_t \in T} \exp\left(-\frac{\|x - x_t\|^2}{2\sigma^2}\right).
\]

To form the displacement field \(\mathbf{D}(x)\), we incorporate both the magnitude and directional aspects of these influences. The direction is given by the vector \((x - x_t)\), pointing away from each tumor voxel \(x_t\). By weighting this direction with the corresponding Gaussian factor and scaling by a proliferation parameter \(\alpha\), we define:
\[
\mathbf{D}(x) = \alpha \sum_{x_t \in T} \exp\left(-\frac{\|x - x_t\|^2}{2\sigma^2}\right) (x - x_t).
\]

Here, \(\alpha\) modulates the overall intensity of the displacement field. Larger \(\alpha\) values correspond to more aggressive tumor growth, resulting in greater displacement of surrounding tissue.

Tumor infiltration into surrounding tissue is simulated by extending the tumor mask beyond its original boundary by a distance \(\phi\) (in cm). Let \(T_0 \subset \Omega\) be the original tumor mask. We first generate a dilated mask \(T_{\phi}\) that encompasses the original tumor plus a peritumoral region of width \(\phi\):
\[
T_{\phi} = \{ x \in \Omega : \mathrm{dist}(x, T_0) \leq \phi \},
\]
where \(\mathrm{dist}(x, T_0)\) is the Euclidean distance from the voxel \(x\) to the nearest voxel in \(T_0\).

This dilated region can be conceptually divided into concentric “rings” or shells, each corresponding to a small radial increment \(\delta r\):
\[
T_{r,\delta r} = \{ x \in \Omega : r \leq \mathrm{dist}(x, T_0) < r+\delta r \}, \quad 0 \leq r < \phi.
\]
By iterating through these rings from the innermost (\(r \approx 0\)) to the outermost (\(r \approx \phi\)), we fill the peritumoral region in a stepwise manner.

For each voxel \(x \in T_{r,\delta r}\), consider a local 3x3 neighborhood \(\mathcal{N}_x\) centered at \(x\):
\[
\mathcal{N}_x = \{ x' \in \Omega : \|x' - x\|_{\infty} \leq 1 \}.
\]
To simulate infiltration, we replace the intensity value of the center voxel \(x\) with a randomly selected intensity from voxels within the same ring \(T_{r,\delta r}\). Formally, if \(I\) denotes the image intensity function, we define:
\[
I_{\mathrm{inf}}(x) = I(x_{r,\delta r}^{*}),
\]
where \(x_{r,\delta r}^{*}\) is a voxel randomly chosen from the set \(\{ x' \in T_{r,\delta r} : x' \in \mathcal{N}_x \}\). By iterating over all voxels in each ring, this process progressively “fills” each ring with intensities from the tumor or previously infiltrated regions, effectively mimicking infiltration spreading outward from the core.

Once all rings up to the dilation boundary \(\phi\) have been processed, we obtain an extended infiltration mask \(I_{\mathrm{inf}}\). However, this discrete ring-based assignment can lead to abrupt transitions. To mitigate this, we apply a Gaussian filter \(\mathcal{G}_{\sigma_I}\) with a chosen standard deviation \(\sigma_I\):
\[
I_{\mathrm{inf,smooth}}(x) = (I_{\mathrm{inf}} * \mathcal{G}_{\sigma_I})(x),
\]
where \(*\) denotes convolution. 

\section{Jacobian Maps}
For the discrete case, the log-transformed determinant of the Jacobian matrix at each voxel \(v_i\) is expressed as follows:

\begin{equation}
\text{JM} = \log\left(\text{det}(\mathbf{J}_i)\right)
\end{equation}

where \(\mathbf{J}_i\) is the Jacobian matrix at voxel \(v_i\), defined as:

\begin{equation}
\mathbf{J}_i = \begin{pmatrix} 
\partial_x D_x & \partial_y D_x & \partial_z D_x \\
\partial_x D_y & \partial_y D_y & \partial_z D_y \\
\partial_x D_z & \partial_y D_z & \partial_z D_z
\end{pmatrix},
\end{equation}

where \(D_x\), \(D_y\), and \(D_z\) are the displacement field components in the \(x\), \(y\), and \(z\) directions, respectively. In the discrete setting, the partial derivatives in \(\mathbf{J}_i\) are approximated using finite differences, 

\begin{equation}
\partial_x D_x \approx \frac{D_x(i+1, j, k) - D_x(i, j, k)}{\Delta x},
\end{equation}

where \(\Delta x\) is the voxel spacing in the \(x\)-direction, and \(v_{i+1} = (i+1, j, k)\) is the neighboring voxel in the \(x\)-direction. 
The TSC is calculated in each voxel of the non-tumour tissue.

\section*{VIII. Tumor Core Masking in Nonlinear Registration}

In our pipeline for computing displacement and compression maps, the tumor core (including necrotic tissue and pathological enhancement) is excluded from the nonlinear registration step via a binary mask. This strategy, known as \emph{cost-function masking}, serves two main purposes:

\begin{itemize}
  \item \textbf{Preventing intensity‐driven artifacts.} During infiltrative growth, cellular invasion produce local intensity changes in the peritumoral region that can mislead the registration algorithm into interpreting these changes as physical tissue shifts. By masking out the tumor area and basing the deformation optimization solely on healthy tissue, we avoid erroneous estimation of the displacement field.
  
  \item \textbf{Ensuring reliable compression measurements.} The Dynamic Infiltration Rate ($\Phi_R$) compares the radial growth of the core tumor with the median compression (Jacobian map) in the peritumoral zone. Without masking, the tumor’s own expansion would be included in the Jacobian calculation, leading to compression overestimation and distortion of $\Phi_R$. The mask guarantees that only deformations induced in healthy tissue by mass effect are measured.
\end{itemize}

\begin{figure}[h!]
  \centering
  \includegraphics[width=0.7\linewidth]{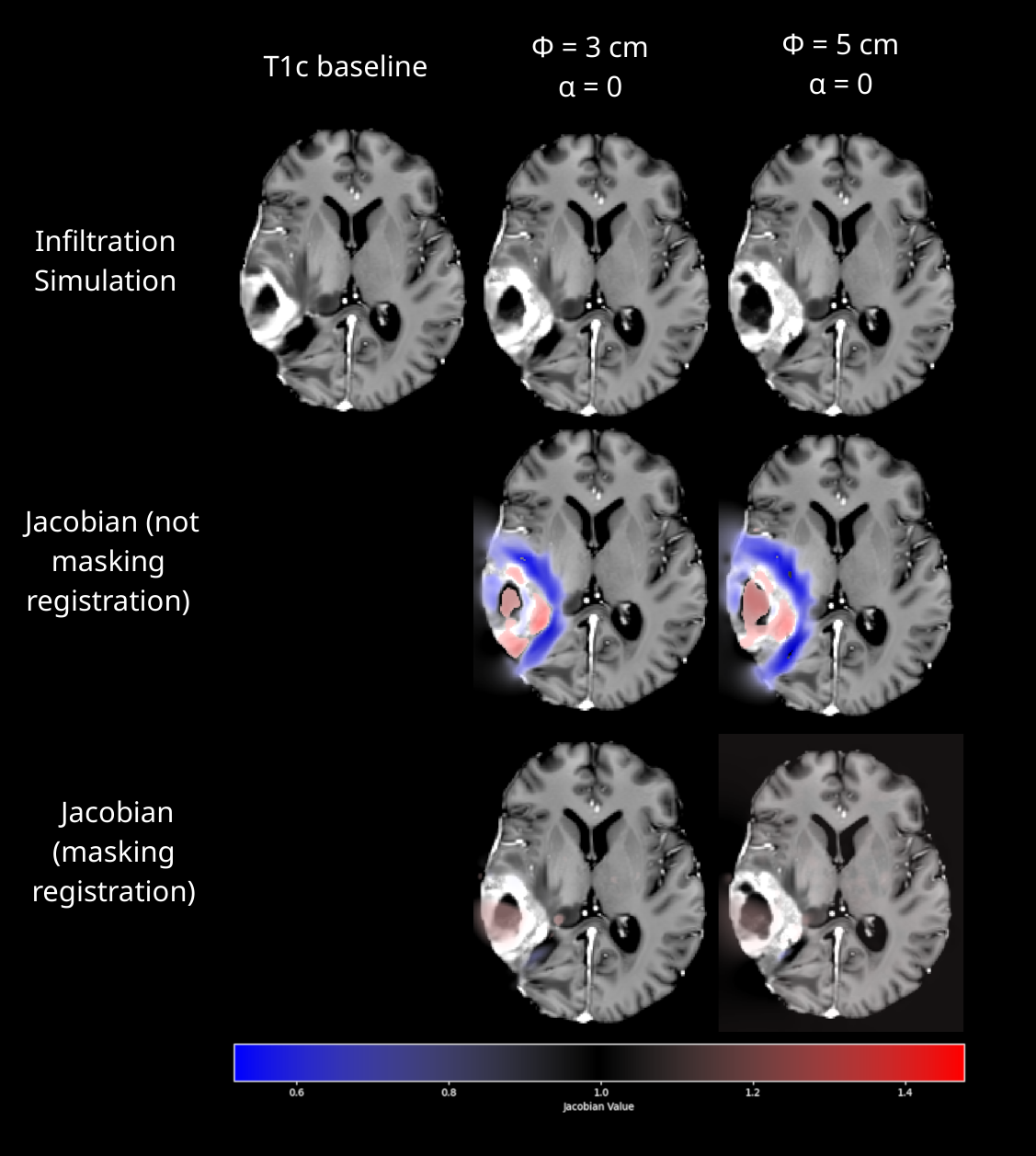}
  \caption{Jacobian determinant maps resulting from deformable registration of the baseline contrast-enhanced T1-weighted (T1c) image under two pure infiltration-only tumor growth simulations (no proliferation, only intensity change), for infiltration distances $\phi = 3cm$ and $\phi= 5cm$. Registration without masking the tumor core: artifactual peritumoral compression ($Jacobian < 1$, indicated by cool colors) appears adjacent to the enhancing tumor margin as the algorithm misinterprets the intensity change as expansion.  Registration with the lesion mask applied: the enhancing core is excluded from the optimization, eliminating false expansion and yielding $Jacobian \approx 1$ (neutral volume change, shown in white) throughout the brain.}
  \label{fig:masking_registration}
\end{figure}

The resulting Jacobian was computed for registration without masking the tumor core and, in parallel, applying the lesion mask for two infiltration-only simulations (3~cm and 5~cm), without proliferation (i.e., no mechanical displacement, only intensity changes) (see Figure \ref{fig:masking_registration}). In both cases the true Jacobian should be zero, since there is no actual compression or expansion—only intensity variation. However, when the tumor core is left unmasked, the registration algorithm falsely interprets the enhancing tumor intensity as expansion and produces peritumoral compression. When the lesion is masked, this artificial compression disappears. The Jacobian in the masked case is much closer to the true value (zero), thereby justifying the use of a lesion-masking cost function in the registration process.






\bibitem{alfonso}
J.~Alfonso, K.~Talkenberger, M.~Seifert, B.~Klink, A.~Hawkins\hyp{}Daarud, and
K.~Swanson \textit{et~al.},
The biology and mathematical modelling of glioma invasion: a review,
\textit{J. R. Soc. Interface} \textbf{14}, 20170490 (2017).
doi:10.1098/rsif.2017.0490

\bibitem{ANTsX_ANTsPy}
B.~B.~Avants, R.~S.~Gill, N.~Tustison, J.~Modat, M.~Robinson, and A.~Rangarajan
\textit{et~al.},
ANTsPy: medical imaging analysis framework in Python,
GitHub repository, \url{https://github.com/ANTsX/ANTsPy}
(accessed 29 Apr 2025).

\bibitem{gerard2017brain}
Gerard, Ian J and Kersten-Oertel et al., 
Brain shift in neuronavigation of brain tumors: A review
Elsevier, Medical image analysis, 35. 03--420 (2017)

\bibitem{biascorrection}
N.~J.~Tustison, B.~B.~Avants, P.~A.~Cook, Y.~Zheng, A.~Egan, and P.~A.~Yushkevich
\textit{et~al.},
N4ITK: improved N3 bias correction,
\textit{IEEE Trans. Med. Imaging} \textbf{29}, 1310–1320 (2010).
doi:10.1109/TMI.2010.2046908

\bibitem{brain_herniation}
C.~A.~Faraj, R.~I.~Snyder, and I.~E.~McCutcheon,
Intracranial emergencies in neurosurgical oncology: pathophysiology and clinical management,
\textit{Emerg. Cancer Care} \textbf{1}, 13 (2022).
doi:10.1186/s44201-022-00013-6

\bibitem{WHO}
Louis, D. N., Perry, A., Wesseling, P., Brat, D. J., Cree, I. A., Figarella-Branger, D., \textit{et al.}, The 2021 WHO classification of tumors of the central nervous system: a summary. \textit{Neuro-oncology}, 23(8), 1231-1251.

\bibitem{california}
B.~K.~Fields, Y.~Xiao, D.~Zhang, A.~G.~Ellingson, J.~C.~Rhodes, and S.~Prado
\textit{et~al.},
The University of California San Francisco Adult Longitudinal Post-Treatment Diffuse Glioma (UCSF-ALPTDG) MRI dataset,
\textit{Radiol. Artif. Intell.} e230182 (2024).
doi:10.1148/ryai.230182

\bibitem{comba2021uncovering}
A.~Comba, S.~M.~Faisal, M.~L.~Varela, T.~Hollon, W.~N.~Al\hyp{}Holou, and
Y.~Umemura \textit{et~al.},
Uncovering spatiotemporal heterogeneity of high-grade gliomas: from disease biology to therapeutic implications,
\textit{Front. Oncol.} \textbf{11}, 703764 (2021).
doi:10.3389/fonc.2021.703764

\bibitem{council}
M.~R.~Council, D.~F.~Smith, J.~Anderson, P.~Goodkin, L.~Jones, and S.~Brown,
Prognostic factors for high-grade malignant glioma: development of a prognostic index. A report of the Medical Research Council Brain Tumour Working Party,
\textit{J. Neuro\hyp{}Oncol.} \textbf{9}, 47–55 (1990).
doi:10.1007/BF00167068

\bibitem{drumm}
M.~R.~Drumm, N.~Andrews, J.~H.~Li, K.~K.~Raghavan, P.~P.~Engell\hyp{}Noerregaard, and
A.~R.~Atkins \textit{et~al.},
Extensive brainstem infiltration, not mass effect, is a common feature of end-stage cerebral glioblastomas,
\textit{Neuro Oncol.} \textbf{22}, 470–479 (2020).
doi:10.1093/neuonc/noz216

\bibitem{gamburg}
E.~S.~Gamburg, W.~F.~Regine, R.~A.~Patchell, J.~M.~Strottmann, M.~Mohiuddin, and
A.~B.~Young,
The prognostic significance of midline shift at presentation on survival in patients with glioblastoma multiforme,
\textit{Int. J. Radiat. Oncol. Biol. Phys.} \textbf{48}, 1359–1362 (2000).
doi:10.1016/S0360-3016(00)00749-9

\bibitem{greedy}
P.~A.~Yushkevich, J.~Pluta, H.~Wang, L.~E.~Wisse, S.~Das, and D.~Wolk,
IC-P-174: fast automatic segmentation of hippocampal subfields and medial temporal lobe subregions in 3 T and 7 T T2-weighted MRI,
\textit{Alzheimers Dement.} \textbf{12}, P126–P127 (2016).
doi:10.1016/j.jalz.2016.06.365

\bibitem{greedy_git}
P.~A.~Yushkevich,
Greedy: fast medical image registration,
GitHub repository, \url{https://github.com/pyushkevich/greedy}
(accessed 29 Apr 2025).

\bibitem{Jain}
R.~K.~Jain, J.~D.~Martin, and T.~Stylianopoulos,
The role of mechanical forces in tumor growth and therapy,
\textit{Annu. Rev. Biomed. Eng.} \textbf{16}, 321–346 (2014).
doi:10.1146/annurev-bioeng-071813-105259

\bibitem{leow2007statistical}
A.~D.~Leow, I.~Yanovsky, M.\hspace{0.2em}C.~Chiang, A.~D.~Lee, A.~D.~Klunder, and
A.~Lu \textit{et~al.},
Statistical properties of Jacobian maps and the realization of unbiased large-deformation nonlinear image registration,
\textit{IEEE Trans. Med. Imaging} \textbf{26}, 822–832 (2007).
doi:10.1109/TMI.2007.892646

\bibitem{merrell}
R.~T.~Merrell, E.~C.~Quant, and P.~Y.~Wen,
Advances in treatment options for high-grade glioma—current status and future perspectives,
\textit{US Neurol.} \textbf{6}, 54–61 (2010).
doi:10.17925/ENR.2010.05.02.54

\bibitem{mgmt}
E.~R.~Gerstner, S.~Yip, D.~Wang, D.~Louis, A.~Iafrate, and T.~Batchelor,
MGMT methylation is a prognostic biomarker in elderly patients with newly diagnosed glioblastoma,
\textit{Neurology} \textbf{73}, 1509–1510 (2009).
doi:10.1212/WNL.0b013e3181c0664b

\bibitem{midline}
J.~Wach, M.~Hamed, P.~Schuss, E.~Güresir, U.~Herrlinger, and H.~Vatter
\textit{et~al.},
Impact of initial midline shift in glioblastoma on survival,
\textit{Neurosurg. Rev.} \textbf{44}, 1401–1409 (2021).
doi:10.1007/s10143-020-01370-8

\bibitem{mirimanoff}
R.\hspace{0.15em}O.~Mirimanoff,
High-grade gliomas: reality and hopes,
\textit{Chin. J. Cancer} \textbf{33}, 1 (2014).
doi:10.5732/cjc.014.10187

\bibitem{paper_elies}
E.~Fuster\hyp{}Garcia, I.~T.~Hovden, S.~F.~Svensson, C.~Larsson, J.~Vardal, and
A.~Bjørnerud \textit{et~al.},
Quantification of tissue compression identifies high-grade glioma patients with reduced survival,
\textit{Cancers} \textbf{14}, 1725 (2022).
doi:10.3390/cancers14071725

\bibitem{seano}
G.~Seano, E.~N.~Henriksson, G.~Gibilisco, C.~G.~Shettigara, B.~He, and
F.~Tremblay\hyp{}Gravel \textit{et~al.},
Solid stress in brain tumours causes neuronal loss and neurological dysfunction and can be reversed by lithium,
\textit{Nat. Biomed. Eng.} \textbf{3}, 230–245 (2019).
doi:10.1038/s41551-018-0326-y

\bibitem{suter2022lumiere}
Y.~Suter, U.~Knecht, W.~Valenzuela, M.~Notter, E.~Hewer, and P.~Schucht
\textit{et~al.},
The LUMIERE dataset: longitudinal glioblastoma MRI with expert RANO evaluation,
\textit{Sci. Data} \textbf{9}, 768 (2022).
doi:10.1038/s41597-022-01644-9

\bibitem{steed}
T.~C.~Steed, J.~M.~Treiber, M.~G.~Brandel, K.~S.~Patel, A.~M.~Dale, and
B.~S.~Carter \textit{et~al.},
Quantification of glioblastoma mass effect by lateral ventricle displacement,
\textit{Sci. Rep.} \textbf{8}, 2827 (2018).
doi:10.1038/s41598-018-20940-1


\begin{thebibliography}{20}

\bibitem{alfonso}
J.~Alfonso, K.~Talkenberger, M.~Seifert, B.~Klink, A.~Hawkins\hyp{}Daarud, and
K.~Swanson \textit{et~al.},
The biology and mathematical modelling of glioma invasion: a review,
\textit{J. R. Soc. Interface} \textbf{14}, 20170490 (2017).
doi:10.1098/rsif.2017.0490

\bibitem{ANTsX_ANTsPy}
B.~B.~Avants, R.~S.~Gill, N.~Tustison, J.~Modat, M.~Robinson, and A.~Rangarajan
\textit{et~al.},
ANTsPy: medical imaging analysis framework in Python,
GitHub repository, \url{https://github.com/ANTsX/ANTsPy}
(accessed 29 Apr 2025).

\bibitem{gerard2017brain}
Gerard, Ian J and Kersten-Oertel et al., 
Brain shift in neuronavigation of brain tumors: A review
Elsevier, Medical image analysis, 35. 03--420 (2017)

\bibitem{biascorrection}
N.~J.~Tustison, B.~B.~Avants, P.~A.~Cook, Y.~Zheng, A.~Egan, and P.~A.~Yushkevich
\textit{et~al.},
N4ITK: improved N3 bias correction,
\textit{IEEE Trans. Med. Imaging} \textbf{29}, 1310–1320 (2010).
doi:10.1109/TMI.2010.2046908

\bibitem{brain_herniation}
C.~A.~Faraj, R.~I.~Snyder, and I.~E.~McCutcheon,
Intracranial emergencies in neurosurgical oncology: pathophysiology and clinical management,
\textit{Emerg. Cancer Care} \textbf{1}, 13 (2022).
doi:10.1186/s44201-022-00013-6

\bibitem{WHO}
Louis, D. N., Perry, A., Wesseling, P., Brat, D. J., Cree, I. A., Figarella-Branger, D., \textit{et al.}, The 2021 WHO classification of tumors of the central nervous system: a summary. \textit{Neuro-oncology}, 23(8), 1231-1251.

\bibitem{california}
B.~K.~Fields, Y.~Xiao, D.~Zhang, A.~G.~Ellingson, J.~C.~Rhodes, and S.~Prado
\textit{et~al.},
The University of California San Francisco Adult Longitudinal Post-Treatment Diffuse Glioma (UCSF-ALPTDG) MRI dataset,
\textit{Radiol. Artif. Intell.} e230182 (2024).
doi:10.1148/ryai.230182

\bibitem{comba2021uncovering}
A.~Comba, S.~M.~Faisal, M.~L.~Varela, T.~Hollon, W.~N.~Al\hyp{}Holou, and
Y.~Umemura \textit{et~al.},
Uncovering spatiotemporal heterogeneity of high-grade gliomas: from disease biology to therapeutic implications,
\textit{Front. Oncol.} \textbf{11}, 703764 (2021).
doi:10.3389/fonc.2021.703764

\bibitem{council}
M.~R.~Council, D.~F.~Smith, J.~Anderson, P.~Goodkin, L.~Jones, and S.~Brown,
Prognostic factors for high-grade malignant glioma: development of a prognostic index. A report of the Medical Research Council Brain Tumour Working Party,
\textit{J. Neuro\hyp{}Oncol.} \textbf{9}, 47–55 (1990).
doi:10.1007/BF00167068

\bibitem{drumm}
M.~R.~Drumm, N.~Andrews, J.~H.~Li, K.~K.~Raghavan, P.~P.~Engell\hyp{}Noerregaard, and
A.~R.~Atkins \textit{et~al.},
Extensive brainstem infiltration, not mass effect, is a common feature of end-stage cerebral glioblastomas,
\textit{Neuro Oncol.} \textbf{22}, 470–479 (2020).
doi:10.1093/neuonc/noz216

\bibitem{gamburg}
E.~S.~Gamburg, W.~F.~Regine, R.~A.~Patchell, J.~M.~Strottmann, M.~Mohiuddin, and
A.~B.~Young,
The prognostic significance of midline shift at presentation on survival in patients with glioblastoma multiforme,
\textit{Int. J. Radiat. Oncol. Biol. Phys.} \textbf{48}, 1359–1362 (2000).
doi:10.1016/S0360-3016(00)00749-9

\bibitem{greedy}
P.~A.~Yushkevich, J.~Pluta, H.~Wang, L.~E.~Wisse, S.~Das, and D.~Wolk,
IC-P-174: fast automatic segmentation of hippocampal subfields and medial temporal lobe subregions in 3 T and 7 T T2-weighted MRI,
\textit{Alzheimers Dement.} \textbf{12}, P126–P127 (2016).
doi:10.1016/j.jalz.2016.06.365

\bibitem{greedy_git}
P.~A.~Yushkevich,
Greedy: fast medical image registration,
GitHub repository, \url{https://github.com/pyushkevich/greedy}
(accessed 29 Apr 2025).

\bibitem{Jain}
R.~K.~Jain, J.~D.~Martin, and T.~Stylianopoulos,
The role of mechanical forces in tumor growth and therapy,
\textit{Annu. Rev. Biomed. Eng.} \textbf{16}, 321–346 (2014).
doi:10.1146/annurev-bioeng-071813-105259

\bibitem{leow2007statistical}
A.~D.~Leow, I.~Yanovsky, M.\hspace{0.2em}C.~Chiang, A.~D.~Lee, A.~D.~Klunder, and
A.~Lu \textit{et~al.},
Statistical properties of Jacobian maps and the realization of unbiased large-deformation nonlinear image registration,
\textit{IEEE Trans. Med. Imaging} \textbf{26}, 822–832 (2007).
doi:10.1109/TMI.2007.892646

\bibitem{merrell}
R.~T.~Merrell, E.~C.~Quant, and P.~Y.~Wen,
Advances in treatment options for high-grade glioma—current status and future perspectives,
\textit{US Neurol.} \textbf{6}, 54–61 (2010).
doi:10.17925/ENR.2010.05.02.54

\bibitem{mgmt}
E.~R.~Gerstner, S.~Yip, D.~Wang, D.~Louis, A.~Iafrate, and T.~Batchelor,
MGMT methylation is a prognostic biomarker in elderly patients with newly diagnosed glioblastoma,
\textit{Neurology} \textbf{73}, 1509–1510 (2009).
doi:10.1212/WNL.0b013e3181c0664b

\bibitem{midline}
J.~Wach, M.~Hamed, P.~Schuss, E.~Güresir, U.~Herrlinger, and H.~Vatter
\textit{et~al.},
Impact of initial midline shift in glioblastoma on survival,
\textit{Neurosurg. Rev.} \textbf{44}, 1401–1409 (2021).
doi:10.1007/s10143-020-01370-8

\bibitem{mirimanoff}
R.\hspace{0.15em}O.~Mirimanoff,
High-grade gliomas: reality and hopes,
\textit{Chin. J. Cancer} \textbf{33}, 1 (2014).
doi:10.5732/cjc.014.10187

\bibitem{paper_elies}
E.~Fuster\hyp{}Garcia, I.~T.~Hovden, S.~F.~Svensson, C.~Larsson, J.~Vardal, and
A.~Bjørnerud \textit{et~al.},
Quantification of tissue compression identifies high-grade glioma patients with reduced survival,
\textit{Cancers} \textbf{14}, 1725 (2022).
doi:10.3390/cancers14071725

\bibitem{seano}
G.~Seano, E.~N.~Henriksson, G.~Gibilisco, C.~G.~Shettigara, B.~He, and
F.~Tremblay\hyp{}Gravel \textit{et~al.},
Solid stress in brain tumours causes neuronal loss and neurological dysfunction and can be reversed by lithium,
\textit{Nat. Biomed. Eng.} \textbf{3}, 230–245 (2019).
doi:10.1038/s41551-018-0326-y

\bibitem{suter2022lumiere}
Y.~Suter, U.~Knecht, W.~Valenzuela, M.~Notter, E.~Hewer, and P.~Schucht
\textit{et~al.},
The LUMIERE dataset: longitudinal glioblastoma MRI with expert RANO evaluation,
\textit{Sci. Data} \textbf{9}, 768 (2022).
doi:10.1038/s41597-022-01644-9

\bibitem{steed}
T.~C.~Steed, J.~M.~Treiber, M.~G.~Brandel, K.~S.~Patel, A.~M.~Dale, and
B.~S.~Carter \textit{et~al.},
Quantification of glioblastoma mass effect by lateral ventricle displacement,
\textit{Sci. Rep.} \textbf{8}, 2827 (2018).
doi:10.1038/s41598-018-20940-1

\end{thebibliography}





\nocite{*}
\end{document}